\documentstyle[11pt,aasms4,rotating]{article}

\def\simlt{\lower.5ex\hbox{$\; \buildrel < \over \sim \;$}}                                    
\def\simgt{\lower.5ex\hbox{$\; \buildrel > \over \sim \;$}}                                    
                                                                          
\def\cm{{\rm\,cm}}

\def\pc{{\rm\,pc}}

\def\gcm3{{\rm\,g\,cm^{-3}}}                                                                   
\def\ncm3{{\rm\,cm^{-3}}}

\def\kev{{\rm\,keV}}                                                                           
\def\mas{{\rm\,mas}}

\def\>{$>$}                                                                                    
\def\<{$<$}

\lefthead{Melia et al.}                                                                        
\righthead{}                                                                                   
                                                                                               
\begin{document}                                                                               
\centerline{Submitted to the Editor of the Astrophysical Journal Letters}                      
\vskip 0.5in          
\title{\bf Is Thermal Expansion Driving the Initial\\ Gas Ejection in NGC 6251 ?} 
                                                                                               
\author{\bf Fulvio Melia$^{1,2,3}$, Siming Liu\altaffilmark{2},                                
and Marco Fatuzzo\altaffilmark{4}}                                                             
                                                                                               
\affil{$^2$Physics Department, The University of Arizona, Tucson, AZ 85721}                    
\affil{$^3$Steward Observatory, The University of Arizona, Tucson, AZ 85721}                   
\affil{$^4$Physics Department, Xavier University, Cincinnati, OH 45207}                        
% Notice that each of these authors has alternate affiliations, which                          
% are identified by the \altaffilmark after each name.  The actual alternate                   
% affiliation information is typeset in footnotes at the bottom of the                         
% first page, and the text itself is specified in \altaffiltext commands.                      
% There is a separate \altaffiltext for each alternate affiliation                             
% indicated above.                                                                             
                                                                                               
\altaffiltext{1}{Sir Thomas Lyle Fellow and Miegunyah Fellow.}                                 
                                                                                               
% The abstract environment prints out the receipt and acceptance dates 
% if they are relevant for the journal style.  For the aasms style, they 
% will print out as horizontal rules for the editorial staff to type 
% on, so long as the author does not include \received and \accepted 
% commands.  This should not be done, since \received and \accepted dates
% are not known to the author.

\begin{abstract}
The relativistic jets in AGNs are probably driven by the action of 
supermassive, spinning black holes.  There is very little direct evidence 
for this, however, since the nuclei of active galaxies are difficult to 
study. This is now changing with new, high-resolution multi-wavelength 
observations of nearby sources, such as Sgr A* at the Galactic 
center, and the nucleus of NGC 6251 (hereafter called NGC 6251*).  In this 
paper, we explore the possibility that the radiative properties of the most 
compact region in NGC 6251* may be understood in the same sense as Sgr A*, 
though with some telling differences that may hint at the nature of jet 
formation.  We show that observations of this object with ASCA, ROSAT, HST 
and VLBI together may be hinting at a picture in which Bondi-Hoyle accretion
from an ambient ionized medium feeds a standard disk accreting at $\sim 
4.0\times 10^{22}$ g s$^{-1}$.  Somewhere near the event horizon, this 
plasma is heated to $>10^{11}$ K, where it radiates via thermal synchrotron 
(producing a radio component) and self-Comptonization (accounting for a 
nonthermal X-ray flux).  This temperature is much greater than its virial 
value and the hot cloud expands at roughly the sound speed ($\sim 0.1c$), 
after which it begins to accelerate on a parsec scale to relativistic 
velocities.  In earlier work, the emission from the extended jet has been 
modeled successfully using nonthermal synchrotron self-Compton processes, 
with a self-absorbed inner core.  In the picture we are developing here, 
the initial ejection of matter is associated with a self-absorbed thermal 
radio component that dominates the core emission on the smallest scales.  
The nonthermal particle distributions responsible for the emission in the 
extended jet are then presumably energized, e.g., via shock acceleration, 
within the expanding, hot gas.  The power associated with this plasma 
represents an accretion efficiency of about $0.54$, requiring dissipation 
in a prograde disk around a rapidly spinning black hole (with spin 
parameter $a\sim 1$).
\end{abstract}                                                                                 
                                                                                               
% The different journals have different requirements for keywords.  The                        
% keywords.apj file, found on aas.org in the pubs/aastex-misc directory,                       
% contains a list of keywords used with the ApJ and Letters.  These are                        
% usually assigned by the editor, but authors may include them in their                        
% manuscripts if they wish.                                                                    
                                                                                               
                                                                                               
% That's it for the front matter.  On to the main body of the paper.                           
% We'll only put in tutorial remarks at the beginning of each section                          
% so you can see entire sections together.                                                     
                                                                                               
% In the first two sections, you should notice the use of the LaTeX \cite                      
% command to identify citations.  The citations are tied to the                                
% reference list via symbolic KEYs.  We have chosen the first three                            
% characters of the first author's name plus the last two numeral of the                       
% year of publication.  The corresponding reference has a \bibitem                             
% command in the reference list below.                                                         
%                                                                                              
% Please see the AASTeX manual for a more complete discussion on how to make                   
% \cite-\bibitem work for you.                                                                 
                                                                                               
\section{Introduction}                                          
                               
The giant elliptical NGC 6251 is host to one of the most spectacular radio
sources in the sky (Waggett, Warner, \& Baldwin 1977), and at a distance
of 106 Mpc (for $H_0=70$ km s$^{-1}$ Mpc$^{-1}$), it is the farthest galaxy
for which the presence of a massive black hole has been inferred on
the basis of (e.g., the kinematic) properties of its core.  ASCA observations of its 
central engine (which we hereafter call NGC 6251*) have suggested the presence
of an extremely broad $6.68$ keV Fe emission line (Turner et al. 1997), 
due to fluorescence and back-scattering in the inner region of 
an optically thick accretion disk (Tanaka et al. 1995), though 
XMM has yet to confirm this. More recent {\it 
Hubble Space Telescope} (HST) optical images of this galaxy's core have 
facilitated a detailed study of its well-defined (and larger) dusty disk, whose Keplerian 
motion points to a central dark mass of $M\sim 4-8\times 10^8\;M_\odot$ 
(Ferrarese \& Ford 1999). This disk, it seems, is about 730 pc in diameter, 
and is inclined by $76^\circ$ to the line of sight.  A significant ionized-gas 
component is confined to the central $\sim 0\farcs3$ ($\approx 150$ pc) region, 
and the fact that the jet is not perpendicular to the dusty disk's plane suggests
that this $730$ parsec-scale structure is misaligned with respect to the
hypothesized subparsec-scale inner accretion disk.

The nucleus of NGC 6251 may be a member of the (relatively small)
class of low-luminosity Active Galactic Nuclei (AGNs), for which
reasonably secure black hole masses have been determined using dynamical
measurements. The faintness of the core emission in these objects makes
them difficult to observe, though several have been studied with
sufficient precision for us to infer that their spectral energy
distributions differ considerably from those of the luminous AGNs. This
class of nuclei includes NGC 1316 and NGC 3998 (Fabbiano, Fassnach, \&
Trinchieri 1994), Sgr A* (Melia 1994;  Narayan, Yi, \& Mahadevan 1995;
Melia \& Falcke 2001), NGC 3031 (M81; Ho, Filippenko, \& Sargent 1996), NGC
4258 (Lasota et al. 1996), M87 (Reynolds et al. 1996), and NGC 4594
(Fabbiano \& Juda 1997), among others (see, e.g., Ho 1999). Of these, Sgr
A* at the Galactic center has been the most extensively discussed source,
in part due to its proximity, which facilitates highly sensitive
measurements with remarkable spatial resolution (now approaching $\sim 1$
A.U., or better; Falcke, Melia, \& Agol 2000).

NGC 6251* appears to have many features in common with
Sgr A*, yet they differ in several significant and important ways. For
example, NGC 6251* is about 200 times more massive than Sgr A*, and is
correspondingly more luminous, but their core spectra are quite similar. On
the other hand, NGC 6251* produces an obvious jet, whereas Sgr A* does
not. Thus, a comparative study of these objects can be valuable in helping
us to understand the underlying physical basis for their activity.  A
principal goal of this paper is to determine if NGC 6251* can be understood 
in the context of the accretion model for Sgr A*, thus helping us 
disentangle the physics of the nascent jet from the underlying processes 
driving the central engine.

It is, in fact, the jet in NGC 6251 that has dominated the study of this
FR I radio galaxy (see, e.g., Bridle \& Perley 1984). Morphologically, 
the jet and its core appear to be resolved into self-similar structures
as one observes this source at progressively higher frequencies, and
therefore with higher spatial resolution (see, e.g., Fig. 12 in
Jones et al. 1986). The radio spectrum of the jet itself is 
adequately described by a nonthermal synchrotron self-Compton model 
(see, e.g., Ghisellini et al. 1993; Guerra \& Daly 1997; Mack et al. 
1997).  The natural interpretation for this phenomenon is that the 
core is associated with optically thick emission at the base of the jet,
probably due to a nonthermal distribution of particles sweeping outward
beyond several parsecs.  However, more recent observations of NGC 6251* 
on even smaller scales (within a fraction
of a parsec) suggest that most of the particle acceleration occurs away 
from the hypothesized black hole in this system (see below).  It is 
therefore possible that the particle distribution is evolving with
distance inside the most compact region resolved thus far. 
                                                         
Our motivation for pursuing this idea stems from our study of Sgr A*, which 
shows that the basic accretion picture for nuclear black holes separates neatly 
into several categories, each of which is characterized by (i) the specific 
angular momentum accreted by the gas at the Bondi-Hoyle capture radius 
($r_{\rm cap}\equiv 2GM/ v_\infty^2$, in terms of the ambient gas 
velocity $v_\infty$), which determines the circularization radius of 
the accretion flow, and (ii) the relative importance of cooling compared 
to compressional heating at $r_{\rm cap}$.  For typical
conditions in the interstellar medium (ISM), the initial temperature
$T[r_{\rm cap}]$ ($\sim 10^6-10^7$ K) sits on the unstable branch of the cooling
function.  Depending on the actual value of $T(r_{\rm cap})$ and the accretion
rate $\dot M$, the plasma settles either onto a hot branch (attaining a
temperature as high as $10^{10} K$ or more at small radii) or a cold branch, in
which $T$ drops to $\sim 10^4$ K.  It would at first appear that NGC 6251* should
be a member of the cold branch family, since the spectrum of its prominent UV emission 
can be fitted very well with a standard, geometrically thin disk, whose existence is
further motivated by the appearance of the broad Fe emission line in its spectrum. 
At higher energies, ROSAT observations suggest that the X-ray emission from NGC
6251* includes two contributions: a resolved thermal component with an
extension of $2\farcs5\sim1.3$ kpc and a temperature of $0.5$ keV, and
an unresolved power-law component that correlates well with the core radio
emission (Worrall \& Birkinshaw, 1994). Even so, the behavior of NGC 6251*
cannot be easily explained in this fashion because 
the inferred accretion rate from the UV spectral fit is $\sim 4.0\times 10^{22}$ 
g s$^{-1}$, which is not consistent with the power produced by the central engine
if the plasma remains cold all the way to the event horizon.
Instead, as we shall see, the strong radio emission from the most compact
region of NGC 6251* may be due to synchrotron emission by a very
hot plasma (in excess of $10^{11}$ K; Jones et al. 1986) within a few thousand 
Schwarzschild radii of the black hole, powering emission at about $0.5$
of the rate with which rest mass energy is carried inward.  Under such conditions, 
the hot plasma would then be unbounded.  

Recent VLBA observations show that the extended radio jet accelerates
on a sub-parsec scale, from $\sim 0.13c$ at $0.30$ pc ($\sim5\times 10^3
r_S$) to $\sim 0.42c$ at $0.57$ pc (Sudou et al. 2000). In this regard,
it is noteworthy that the temperature ($\sim 10^{11}$ K) required to 
produce the strong radio emission from the most compact region measured 
to date (assuming a compact thermal particle distribution) corresponds to 
a sound speed of $\sim 0.1\ c$. The coincidence of these speeds strongly 
suggests to us that the nascent jet may be formed by the same hot expanding 
plasma that would then produce the VLBA core component via thermal synchrotron 
radiation. If this picture is correct, the nonthermal particles inferred from the 
jet emission on larger scales would then be accelerated---presumably via
shock acceleration or an electrodynamic process---within the expanding hot gas. 
In this paper, we scrutinize the multi-wavelength observations
of NGC 6251*, and invoke a hot expanding plasma to fit the radio spectrum of
the nucleus and its self-Comptonized component, which in turn appears to account 
well for the unresolved power-law X-rays.  We shall find that the implied mass 
loss rate is a significant fraction of the accretion rate through the disk.

\section{Multi-Wavelength Observations of the Nucleus in NGC 6251} 

The currently available observations of NGC 6251* are summarized in Table 1. 
VLBI maps of NGC 6251* have sufficient resolution to separate out the jet emission 
from the core down to a size of about $2,500$ Schwarzschild radii (e.g., Cohen 
\& Readhead 1979;  Jones et al. 1986); the corresponding flux varies significantly 
on a time scale of years. It should be emphasized that the cm-mm radio spectrum 
of the core on this scale differs considerably from that of the extended and 
the sub-parsec scale jet components.  Whether thermal, nonthermal, its rise
toward mm wavelengths---reminiscent of the so-called mm to sub-mm bump in Sgr A* 
(Melia \& Falcke 2001)---is a signature of self-absorbed synchrotron emission
(e.g., Melia 1992, 1994).  The possibility that the core component on this smallest
scale is associated with the jet dynamics is underscored by the recent
discovery of a sub-pc-scale counter jet in this source (Sudou et al.
2000), which shows that the outflow is accelerated from $\sim 0.13c$ at
$0.30$ pc to $\sim 0.42c$ at $0.57$ pc.  The nascent jet therefore appears to
accelerate to its terminal velocity well beyond the compact inner region studied here.

At X-ray energies, Birkinshaw \& Worrall (1993)
analyzed the ROSAT/PSPC data and concluded that nearly all ($\sim
90\%$) of the emission arises from a spatially unresolved power-law component
with a diameter $< 200$ pc ($\sim 3.5\times 10^6 r_S$);  this component
correlates well with the core radio emission (Worrall \& Birkinshaw 1994).  
Data now exist for the core of NGC 6251 across at least 9 orders of
magnitude in frequency (see Fig. 1 below).  The {\it ROSAT} observations are
important also because they place a lower limit (of $\sim 0.1$ cm$^{-3}$) on
the proton density of the ISM and the inferred intrinsic hydrogen column
density is $5.0\times 10^{20}\cm$. Ferrarese \& Ford (1999) argue that the
implied Bondi-Hoyle accretion rate is ($\sim 5\times
10^{-4}\,M_\odot$ yr$^{-1}\approx 3.2\times 10^{22}$ g s$^{-1}$).

\small{
\begin{quote}
\begin{tabular}{ccccccc}
\hline
\hline
$\lambda$       & $\nu$  & $F_\nu$ &Integration & Telescope &
Date & Notes\\
or Energy band    &  (Hz)  &   (Jy)  &Area      & or Instrument&      &
\\
\hline
$13$ cm  & $2.3\times 10^9$ & $0.65$ & $0.5\mas\times 0.6\mas$ & VLBI & May
1978 & 1 \\
$2.8$ cm  & $1.1\times 10^{10}$ & $0.9$ & $0.5\mas\times 0.6\mas$ & VLBI & May
1978 & 1 \\
$18$ cm& $1.7\times 10^9$ & $0.28$ & $3.0\mas\times 3.0\mas$ & VLBI & Mar 1983
& 2\\
$13$ cm & $2.3\times 10^9$ & $0.25$  & $2.5\mas\times
1.8\mas$ & VLBI & Feb 1980 & 2\\
$6$ cm & $5.0\times 10^9$ & $0.35$ & $1.3\mas\times
1.1\mas$ & VLBI & Dec 1981 & 2\\
$6$ cm & $5.0\times 10^9$ & $0.13$ & $0.5\mas\times
0.5\mas$ & VLBA & Apr 1998 & 3\\
$2$ cm & $1.5\times 10^{10}$ & 0.34 & $0.5\mas\times
0.5\mas$ & VLBA & Jun 1998 & 3 \\
$0.81\,\mu$m & $3.69\times 10^{14}$ & $3.7\times10^{-4\dagger}$ &
$0\farcs1\times0\farcs1$ & WFPC2 on HST & Jun 1995 & 4 \\
$0.56\,\mu$m & $5.41\times 10^{14}$ & $2.7\times10^{-4\dagger}$ &
$0\farcs1\times0\farcs1$ & WFPC2 on HST & Jun 1995 & 4 \\
$0.41\,\mu$m & $7.32\times 10^{14}$ & $1.6\times10^{-4\dagger\ddagger}$ &
$0\farcs1\times0\farcs1$ & FOC on HST & Feb 1996 & 4 \\
$0.34\,\mu$m & $8.82\times 10^{14}$ & $9.6\times10^{-5\dagger\ddagger}$ &
$0\farcs1\times0\farcs1$ & FOC on HST & Feb 1996 & 4 \\
0.2-2.4 keV  &$(0.5-5.8)\times 10^{17}$ & $3.7\times 10^{-7*}$ &
$4\farcs0 \times 4\farcs0$ & ROSAT & Mar 1991 & 5\\
\hline 
\end{tabular}

\hspace{0.5cm}Notes: (1) Cohen, \& Readhead 1979. (2) Jones et
al. 1986. (3) Sudou et al. 2000. (4) Crane \& Vernet 1997. (5) Birkinshaw \&
Worrall 1993.

\hspace{0.5cm} $\dagger$ Adopting $A_v= 0.88$ mag (Ferrarese \& Ford 1999).\newline
$\ddagger$ The intrinsic flux may be larger than the value given here by $10\%$ due to
the unknown nonlinearities in FOC (Crane \& Vernet 1997).\newline
* This is the flux density at $1\kev$, and the spectral index of the 
power-law component is $1.0\pm0.5$.
\end{quote}
}

\normalsize

Another important observation was made by ASCA. Its high spectral
resolution makes it possible to resolve the emission line in the X-ray
band. In NGC 6251*, the equivalent width of the Fe xxv $6.68$ keV 
line is $392\pm305$ eV (Turner et al. 1997). This
extreme width corresponds to a velocity of $\sim0.3 c$, reminiscent
of the accretion disk model for MCG-6-30-15 (Tanaka et al. 1995).

\section{Structure of the Emitting Gas around NGC 6251*}

Roughly speaking, the following inequality should be satisfied if the emitting
gas near the central black hole is bounded:
\begin{equation}
\alpha\, k_b \,T\leq \left(1-\sqrt{1-{r_S\over r}}\right)m_p c^2\ ,
\end{equation}
where $\alpha=3$ for a fully ionized but non-relativistic plasma, and
$\alpha=9/2$ when the electrons are relativistic. Also, $k_b$ is the Boltzmann 
constant, $m_p$ is the proton mass, and $T$ is the temperature of the (assumed
thermal) plasma. For NGC 6251*, the Schwarzschild radius is $r_S=1.77\times 
10^{14}\cm$ (assuming a mass of $6\times 10^8\,M_\odot$). Thus, we should have 
\begin{equation}
T\leq 3.6\times 10^{12}\left(1-\sqrt{1-{r_S\over r}}\right) \;\hbox{K}\ ,
\label{ineq1}
\end{equation}
if the emitting gas is bounded.

However, this is not consistent with observations of NGC 6251* at $5$ GHz, 
which show that the flux density of the core is $0.13$ Jy$/$beam, with a
beam size of $0.5\mas\times0.5\mas$. Because this flux should be less than
that produced by an optically thick plasma, we have 
\begin{equation} 
F_\nu\leq \left({R\over D}\right)^2 {2\pi\, k_b\, T\nu^2\over c^2}\ ,
\end{equation} 
where $F_\nu$ is the flux density, $R$ is the characteristic radius of the 
emitting region, and $D=106$ Mpc is the distance to NGC 6251. That is,
$T({R/ r_S})^2\geq 1.8\times 10^{17} \;\hbox{K}$. Thus, if the radio emission 
is produced by a bound plasma, $R/r_S\geq 2.3\times 10^5$, 
corresponding to an angular size of $25\mas$, which is much larger
than one half the beam size within which the radio emission is supposed to be
confined. Thus, if the gas producing the radio emission in NGC 6251* is
thermal, it must have a temperature above its virial value.
More specifically, the fact that the radio emission is confined to
a region apparently no larger than $0.25\mas$ ($\sim 2,400 \,r_S$),
the required temperature is $T>3.3\times
10^{10}$ K.  Here, we have ignored the effects of Doppler boosting, justified 
on the basis that the bulk velocity of the radio-emitting plasma is about 
$0.1\,c$. 

The UV spectrum and the tentative observation of a broad Fe emission line also 
suggest the presence of a standard optically thick disk, which must merge into
the hot plasma described above. The size of this disk is determined by
the specific angular momentum accreted with the gas following Bondi-Hoyle 
capture.  Its temperature is a measure of the accretion rate.  Since
most of the UV radiation is produced near the inner region of this
circularized flow, only the temperature (and hence the accretion rate)
is an important adjustable parameter for fitting the observations.

On an even larger scale, the HST observations indicate that there exists a
significant ionized-gas component within $150\pc$ from the central black
hole. For an accretor with mass $\sim 6\times 10^8\;M_\odot$, the capture
radius is approximately $20$ pc (scaled to an ambient gas velocity of
$500$ km s$^{-1}$), so this ionized-gas component must contribute
a substantial fraction of the Bondi-Hoyle captured gas that flows inward
to merge with the accretion disk as smaller radii (see below).  The ROSAT observation
shows that this gas component should also contribute to the extended thermal
X-ray source, not unlike the situation now known to exist at the Galactic
center following imaging observations with {\it Chandra}
(e.g., Baganoff et al. 2001). The larger scale dusty disk imaged by
HST has a radius of about $250\pc$, and since this structure is relatively
faint, we will not include it in our spectral modeling.
 
\section{Calculation of the Spectrum}

There are three dominant spectral components in our model: one is the
thermal synchrotron (radio) emission from the expanding hot nuclear
region---which we assume dominates the observed core radio flux associated
with the inner $2,500$ Schwarzschild radii; the second is the self-Comptonized 
radiation from this region (contributing mostly to the nonthermal X-ray flux); 
and the third is the optically thick emission from the standard accretion disk. 
The latter is a function of the accretion rate and the inclination angle. Based 
on their best fit to the line emission from the nucleus, Ferrarese and Ford (1999) 
inferred an inclination angle of $31^\circ$. Fitting the UV data with such a disk 
(see Figure 1), we find that the required accretion rate is $\sim 4.0\times 
10^{22}$ g s$^{-1}$. This is significant in view of the fact that the Bondi-Hoyle 
capture rate within the ionized-gas component is very close to this value 
(Ferrarese and Ford 1999).  It seems, therefore, that the disk is fed almost 
entirely from the ambient ionized medium, and that any energy flux advected through
the disk must be a small fraction of the total power (cf. Narayan et al. 1995).

To model the thermal synchrotron-emitting plasma, it is necessary to make some 
simplifying assumptions, since we do not yet know precisely how this region is
energized (but see Krolik 1999, Gammie 1999, and Agol \& Krolik 2000 for some 
recent work on this topic).  The structure (e.g., the radial stratification) of 
the inner hot plasma must therefore reflect the nature of this heating mechanism.
As a first step, we will simply take this inner region to be a uniform
sphere.  Calculating the thermal synchrotron spectrum and its self-Comptonized
component then produces the best fit model shown in Figure 1---always under the
assumption that any self-absorbed nonthermal synchrotron emission on this scale
is small by comparison.  The inset shows an enlarged view of the fit to the radio 
spectrum of the core in NGC 6251, together with data gathered at three different epochs 
(each set being connected by dashed lines). Our fit corresponds to the most recent
observation, represented by the two lowest points in the diagram, since this
occurred closest to the observations at other wavelengths. The calculated 
radio spectral index is somewhat larger than that observed, but given the
oversimplified geometry adopted for this calculation, and the strong temporal
variability in the radio spectrum, the comparison is rather promising.

The best fit model for the hot, expanding gas has a radius $R=450 r_S$,
a temperature $T= 10^{12}$ K, an electron number density $n_e=300$
cm$^{-3}$, and a magnetic field $B=0.06$ G. All these parameters are guided
by, and are consistent with, the limits set by Jones et al. (1986).
As one can see from this figure, the radio spectrum produced in the optically
thick region is that of a blackbody, consistent with a spectral index of 2. 
In the optically thin region, the measured flux density is given by
$F_\nu=\epsilon_\nu V/4\pi D^2$, where $V$ is the volume of the sphere.
The emissivity $\epsilon_\nu$ for thermal
synchrotron emission is well-known and is given in Pacholczyk (1970). 
Also, although we here calculate the Comptonized component numerically, it
is straightforward to estimate it as follows. At a temperature of $10^{12}$ K, 
the characteristic particle Lorentz factor is $\gamma_e\simeq 3k_bT/m_e
c^2=506$. Then Compton-scattering will increase the photon frequency by a
factor $\sim\gamma_e^2=2.6\times 10^5$. In reality, the electron Maxwell-Boltzmann
distribution broadens the Comptonized component somewhat. The ratio of the 
scattered peak flux density to that in the radio is therefore  $\sim\sigma_T 
R n_e$, which is roughly $10^{-5}$ for the best fit parameters. These 
values are in excellent agreement with the results of numerical calculations 
shown in Figure 1. In particular, it is evident from the figure that this compact 
nucleus can fit the radio and X-ray spectral components self-consistently. 
Indeed, because the X-ray emission is produced by self-Comptonization of the 
radio photons, these two components must be highly correlated, consistent 
with the assessment made by Worrall and Birkinshaw (1994).

The UV flux is produced predominantly within the inner $\sim 20r_S$ of the
optically thick accretion disk, whose inner radius in this simulation is set
at $3r_S$. Note that if the black hole is spinning, this inner radius is
reduced even further (for a prograde orbit), making the required accretion rate 
somewhat smaller than the $\sim 4.0\times 10^{22}$ g s$^{-1}$ inferred above. 
The spectral fit in this waveband appears to be quite satisfactory also. 

Notice, however, that at EUV energies, both the disk and the self-Comptonized
component from the expanding plasma contribute to the spectrum.  The
latter dominates progressively more and more toward higher frequency.  The
fact that the Comptonized component is produced in a relatively large region 
may explain the observed asymmetry of the EUV source. The HST observations found 
that only one side of the large scale dusty disk is illuminated, suggesting a 
warped geometry (Crane and Vernet, 1997). However, in the UV band, such a reflection 
component does not exist, indicating that the UV and EUV emissions from the inner
region might have different origins. According to our model, the UV emission
is produced in the inner region of the small accretion disk, whereas the EUV
radiation is due to self-Comptonization within the much larger expanding hot
cloud. In this picture, it is therefore much easier to illuminate the large
dusty disk at EUV rather than UV energies, which appears to be borne out by
the observations.

\section{Conclusions}

>From the parameters of the best fit model, we can estimate the rate of
mass loss associated with the expanding hot plasma in the nucleus:
\begin{equation}
\dot{M}_{\rm loss}\sim \Delta\Omega\,r_o^2\,(fc_S)\,m_p\,n\sim 9.6\,
\Delta\Omega\,(fc_S/0.1c)\times 10^{21}\;{\rm g}\;
{\rm s}^{-1}\;,
\end{equation}
where $fc_S$ is observed to be about $0.13c$ at $0.30$ pc, so $f$ may be less
than one closer to the black hole. Also, $\Delta\Omega$ reflects the unknown 
geometry of the expanding gas---the more collimated the expansion is, the smaller 
the value of $\Delta\Omega$ will be, though observationally it appears that
$\Delta\Omega\sim O(1)$. The fact that $\dot{M}_{\rm loss}$ in this picture is 
similar to the inferred accretion rate through the disk suggests that much of the 
infalling matter is energized by the black hole and escapes via thermal expansion.

The need for a large efficiency in converting accreted rest mass energy into the
power of the central source can be inferred on the basis of several lines of evidence.
First, in the X-ray band ($0.2-2.4$ keV), the luminosity is about $1.9\times 10^{42}$ 
ergs s$^{-1}$ (assuming a spectral index of $1.0$, as indicated by our best fit model). 
The total X-ray luminosity is even larger than this. However, the dissipation of 
gravitational energy in the disk accounts for only $2.6\times 10^{42}$ ergs s$^{-1}$, 
which is in fact equal to the UV luminosity. Second, the energy advection rate in the 
expanding cloud would have to be $\sim 1.5\times 10^{43}$ ergs s$^{-1}$ if all of the 
accreting gas (i.e., about $4.0\times 10^{22}$ g s$^{-1}$) is expelled from the nucleus.  
Thus, the total power produced near the black hole is $\sim (15+4.5)\times 10^{42}$ ergs 
s$^{-1}$, which then implies an efficiency $\epsilon\approx 19.5/36\approx 0.54$.  
According to the classical estimates for the accretion efficiency (see, e.g., Bardeen 
1970; Thorne 1974; Abramowicz, Jaroszynski \& Sikora 1978) this inferred value of 
$\epsilon$ requires a rapidly spinning black hole with a spin parameter $a\sim 1$.
A value of $\epsilon > 0.42$ may require additional physics associated with
energy extraction via strong magnetic fields that couple the tightest orbits to those
well beyond the inner edge of the disk (see, e.g., Krolik 1999; Gammie 1999). 

An alternative model, in which no such conversion occurs, but rather the disk is 
advection dominated and thereby releases its trapped energy at small radii, is 
unlikely to apply here.  Given that a large fraction (perhaps all) of the accreted 
mass is lost in the expanding hot cloud, it is difficult to see why this unbound 
plasma would accrete down to small radii in the first place.  In addition, a 
truncated standard disk (Quataert et al. 1999), as introduced in the ADAF model, 
cannot fit the UV spectrum. The reason is that to produce significant EUV emission, 
the temperature of the disk should be around $7,000$ K. For a truncated disk with 
such a temperature, the UV flux density is much larger. Also the properties of the 
ionized gas surrounding NGC 6251* inferred from the ROSAT observations constrain 
the accretion rate rather tightly. A model with a much larger accretion rate is 
not favored. 

In summary, the multi-wavelength observations of NGC 6251*, and the associated
theoretical modeling of this source, point to the following scenario as a viable
description for the initial ejection of hot plasma near the black hole, and
its corresponding radio spectrum, which may dominate the core emission seen 
on the smallest spatial scales. Bondi-Hoyle accretion (starting at $\sim 20$ 
pc) from the surrounding ionized medium feeds a standard disk at a rate of 
$\sim 4.0\times 10^{22}$ g s$^{-1}$.  This disk produces the UV spectral 
component measured by HST, but close to the event horizon (probably near the
innermost stable orbit in a Kerr metric with $a\sim 1$), the infalling gas is 
energized and radiates via thermal synchrotron processes to produce (much, or all, 
of) the core VLBI radio flux.  The temperature of the gas is much greater than its 
virial value, and so the hot plasma expands at $\sim 0.1c$, close to its speed 
of sound.  Very importantly (particularly for future modeling), most of the 
acceleration that produces the large-scale jet therefore does not occur within 
the core, but rather acts some distance away.  In this system, at least, the 
black hole energizes the infalling plasma and expels it to feed the nascent 
jet, but the acceleration that drives the outflow to relativistic velocities 
occurs on a parsec scale, some $10^4\,r_S$ beyond the event horizon.  

{\bf Acknowledgments} We are indebted to the referee for a careful reading
of the manuscript and very thoughtful suggestions for improvement.  This 
research was partially supported by NASA under grants NAG5-8239 and NAG5-9205, 
and has made use of NASA's Astrophysics Data System Abstract Service.  FM is 
very grateful to the University of Melbourne for its support (through a 
Miegunyah Fellowship) and MF thanks the John Hauck Foundation for partial 
support.
                                                                                               
{}
                                                                                               
% And finally, we must deal with the figures.  There are three figures 
% associated with this manuscript; two figures are Encapsulated 
% PostScript (EPS) files.  The third figure is a grey scale figure that does 
% not exist in EPS form.  
% Authors have three options for including figure information within a 
% manuscript.  Not all the options may be acceptable by the target Journal - be 
% sure to look at the appropriate submission instructions, electronic or 
% otherwise.  
% % Option 1.  Using this option, only the figure captions are included in the
% main body of the manuscript.  The figure captions must start on a new page.  
% The captions are generated with the \figcaption[]{} command: the first 
% argument is optional, if you put something in there, put the name of the 
% EPS file that goes with the caption; the second argument is the figure 
% caption itself, and may include a \label command.  The \figcaption command 
% generates the figure numbers.  This option is acceptable for all manuscript
% submissions.  \newpage

\begin{figure}[thb]\label{fig1.ps}                 
{\begin{turn}{0}                                                                               
\epsscale{0.8}                                                                                 
\centerline{\plotone{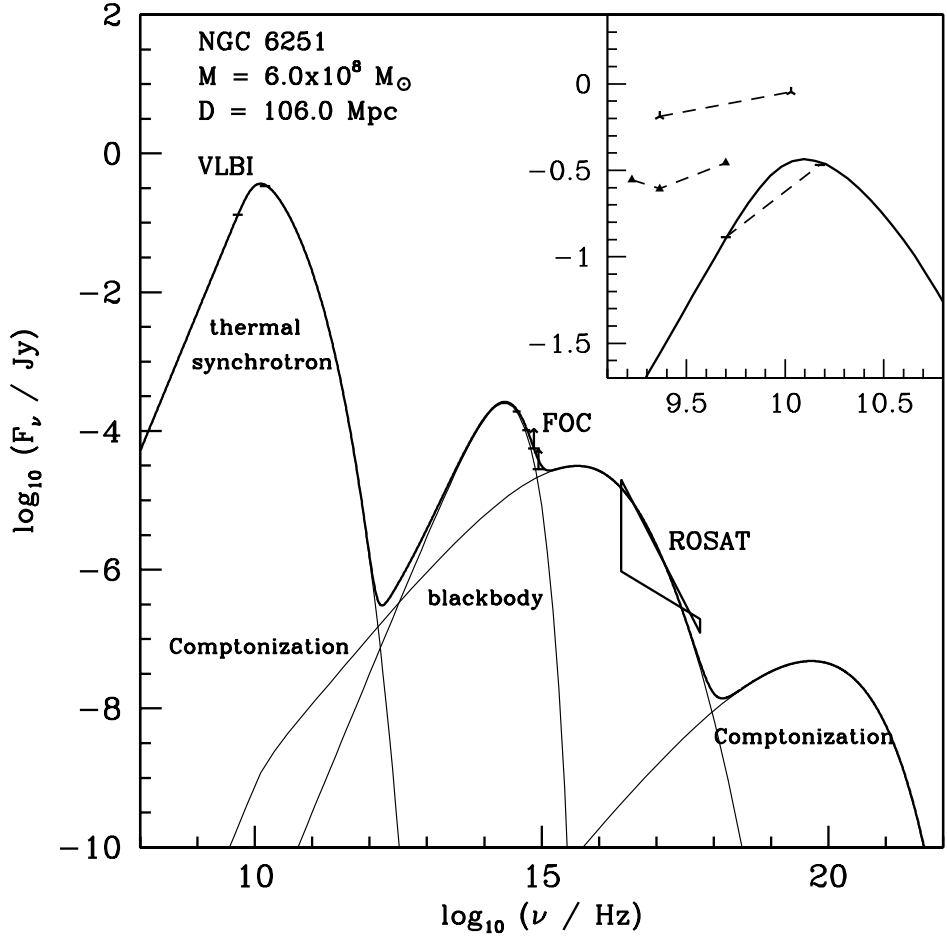}}  
\end{turn}}                                                                                    
\caption{The best fit spectrum compared to the available data for the                          
nucleus of NGC 6251.  The references are given in the text.  The inset                         
shows an enlarged view of the fit to the most recent core spectrum, contrasting                            
with the flux density observed in 1978 (top dashed                          
curve) and that in the earlier 1980s (middle dashed curve) (from Jones                        
et al. 1986).  The solid curve shows the total spectrum from the core.                         
The thermal synchrotron and self-Comptonized components arise in the                           
expanding hot plasma, whereas the blackbody component,                                
accounting for the IR-optical `bump', is due to emission in the inner                           
optically thick, cold disk.}                                                    
\end{figure}                                                                                   
                                                                                               
\end{document}